\documentclass[conference]{IEEEtran}
\IEEEoverridecommandlockouts
\usepackage{amsmath,graphicx,booktabs}

\usepackage{cite}
\usepackage{amsmath,amssymb,amsfonts}
\usepackage{algorithmic}
\usepackage{graphicx}
\usepackage{textcomp}
\usepackage{epstopdf}
\usepackage{setspace}
\usepackage{amsmath}
\usepackage{amssymb}
\usepackage{stfloats}
\usepackage{amsthm}
\usepackage{amsfonts}
\usepackage{color}
\usepackage{graphicx}
\usepackage{bbm}
\usepackage{subfigure}  

\usepackage[flushleft]{threeparttable} 
\usepackage{upgreek}
\usepackage{bm}
\usepackage{nicefrac}

\usepackage{multirow}
\usepackage{cite}
\usepackage{mathtools}
\usepackage{stmaryrd}
\usepackage[mathscr]{euscript}
\usepackage{array}
\usepackage{mathrsfs}

\usepackage{soul}

\usepackage{float}
\usepackage{setspace}
\usepackage{hyperref}
\usepackage{url}
\usepackage{breakurl}

\begin{document}

\title{Expressive Timing in Hindustani Vocal Music}

\author{\IEEEauthorblockN{Yash Bhake and Preeti Rao}
\IEEEauthorblockA{\textit{Department of Electrical Engineering,  Indian Institute of Technology Bombay},
Mumbai, India\\
Email: 22b2148@iitb.ac.in, prao@iitb.ac.in}}

\maketitle

\begin{abstract}

Temporal dynamics are among the cues to expressiveness in music performance in different cultures. In the case of Hindustani music, it is well known that expert vocalists often take liberties with the beat, intentionally not aligning their singing precisely with the relatively steady beat provided by the accompanying tabla. This becomes evident when comparing performances of the same composition such as a \textit{bandish}. We present a methodology for the quantitative study of differences across performed pieces using computational techniques. This is applied to small study of two performances of a popular \textit{bandish} in \textit{raga Yaman}, to demonstrate how we can effectively capture the nuances of timing variations that bring out stylistic constraints along with the individual signature of a performer. This work articulates an important step towards the broader goals of music analysis and generative modelling for Indian classical music performance.

\end{abstract}
%

\section{Introduction}

\textit{Bandish} are notated lyrical compositions in the Hindustani music genre that serve as a reference for the associated \textit{raga} in terms of the overall pitch movement and characteristic phrases. A \textit{bandish} comprises two verses, typically with 2-4 lines each. Vocal concerts include the singing of a \textit{bandish} in the chosen \textit{raga}. Performers bring in different variations in the rendition of a given \textit{bandish} and we also see variations in the same \textit{bandish} lines across repetitions within the performance. The variations are evident to listeners, who are usually familiar with the prototypical form of the \textit{bandish}. The associated auditory experience is greatly appreciated especially when executed by expert musicians. 

In a work that addresses ‘musical expressiveness’ across cultures, Fabian et al.  \cite{fabian} refer to the effects caused by the variation of auditory parameters such as loudness, intensity, phrasing and tempo away from a prototypical performance but strictly within stylistic constraints.  They make it clear that expressiveness does not refer to any features of the composition itself or the emotion that is expressed. The vocal rendition of a \textit{bandish} in a Hindustani music concert can therefore be considered suitable for the investigation of the kind of variation that constitutes musical expressiveness. 

We refer to the widely regarded book of V. N. Bhatkhande \cite{Bhatkhande} who collected and notated a large number of traditional compositions from across the country in the early $20^{th}$ century. The Bhatkhande notated compositions serve as a convenient reference for a discussion of the measured variations of a given \textit{bandish} line across performers. With schematic notation that depicts the melodic outline using the syllables of the lyrics, it is possible to relate the sung performance to the canonical version via the corresponding lyrics syllables.   In this work, we use computational techniques to compare the performances of a given \textit{bandish} by different artists. We present our methodology that includes automatic techniques for some of the audio feature extraction. We focus solely on timing expressiveness, leaving the study of pitch inflections and other dynamics that can also define expressiveness, to future work. In particular, we are interested in developing computational methods that capture the similarities and differences between different performances of the same \textit{bandish}. We illustrate our approach with a small study of audio recordings by two expert vocalists of a well-known \textit{bandish} in \textit{raga Yaman} to draw insights about individual differences as well as the adherance to any genre-specific contraints.  

The motivation for our work is to model the variations in acoustic parameters that can lead to the deeper understanding of music performance and the strategies used by accomplished musicians. This can potentially contribute to quantitative models for music generation from notation that are consistent with the intricacies of the specific genre and style.


\section{Dataset and Preparation}

Table ~\ref{tab:dataset} describes our dataset comprising audio and metadata for several performances of \textit{bandish} in \textit{raga Yaman}, a popular \textit{raga} which is taught early in music training and widely performed on the concert stage.

In terms of choice of \textit{bandish}, we restrict ourselves to \textit{madhyalay} and \textit{drut} (medium and fast tempo) due to the relatively high complexity and freedom in \textit{vilambit} (slow tempo) \textit{bandish}. We select \textit{bandish} where we have the canonical notation from Kramik Pustak Malika \cite{Bhatkhande}. Next, for each \textit{bandish}, we look for good quality audios recordings ranging from maestro concerts to simpler teaching websites to capture performance diversity, including also recordings from Samarpan by Pt. I. Nirody \cite{Nirody}, following the methodology of Madhumitha \cite{madhumitha}. 

The audio performance comprises the singing voice accompanied by the tabla supplying the relatively steady beat and defining the rhythmic grid for the chosen tala. A complete acoustic characterisation in terms of musically relevant parameters would need the onset instants of each of the tabla strokes and the sung syllables, and the notes (including pitch inflections or melodic ornamentation). Given the correspondence between the lyric syllables and the notes in the schematic notation of the composition,  we expect the timing expressiveness to be manifested in the deviation of the sung syllables from the specified beat locations. The reliable detection of onsets of sung syllables apart from tabla stroke onsets is a component of the automated audio processing. It is necessary also to establish a temporal alignment of the note events between the performances to be compared. This is achieved using the matching of the lyrics or syllable sequences. In this section, we present our manually labeled dataset that facilitates the development of the needed text processing as well as audio processing for onsets. 


\vspace{-5pt}
\begin{table}[h!]
\centering
\footnotesize
\begin{tabular}{lcc}
\toprule
\textit{Raga} & \textit{Raga Yaman} \\
\midrule
\# \textit{bandish} & 5  \\
\# Recordings & 14  \\
\# Talas & 1  \\
\# Artists & 8  \\
\# Canonical lines & 33 \\
\# Syllable Onsets & 2041 \\
\# Syllables (canonical) & 274 \\
Matra per min range & 110-180 \\
Total duration (min) &  25.25\\
\bottomrule
\end{tabular}
\vspace{3pt}
\caption{Summary of the dataset used for onset detection evaluation}
\label{tab:dataset}
\end{table}
\vspace{-20pt}

\subsection{Concert audio segmentation}

Our dataset\footnote{The audio files can be accessed in the \href{https://www.notion.so/Supplementary-Material-Expressive-Timing-in-Hindustani-Vocal-Music 137196833673806d9be2cd4c597a0e8d\#137196833673805b9636e036bcae03ec}{supplementary material} \cite{author2025supplementary}}
 is largely comprised of recordings from concerts on YouTube aside from the above-mentioned organized collections. We required good-quality audio with clearly intelligible vocals to facilitate our intended audio processing steps. 
 
Given our interest in the performance of the lines of the \textit{bandish}, we extract these segments from the full recital, doing away with the improvisation fillers such as the \textit{alap} and \textit{vistar}. Source separation with a commercially available tool - Gaudiolab\cite{gaudiolab} was applied to obtain the vocals-only audio and the complementary track with tabla accompaniment, harmonium and the drone.  
The vocal segments are manually marked for syllable onsets, which are further labelled by the syllable as heard (and also consistent with the lyrics of the composition). The manual detection of onsets is greatly facilitated by using a wideband spectrogram display where the consonant-vowel transitions are relatively prominent \cite{praat_software}.


Similarly, for the accompaniment track, we manually annotated the \textit{sam} (1\textsuperscript{st} beat) and \textit{khali} (9\textsuperscript{th} beat), the two salient beats in the applicable 16-beat rhythmic cycle \textit{teentaal}. The intervals between these salient beats were divided into eight equal parts, given that each audio clip in our dataset has a nearly constant tempo as also realised by the \textit{tabla} strokes.


\subsection{Canonical notation}

\begin{figure}
    \centering
    \includegraphics[width=1\linewidth]{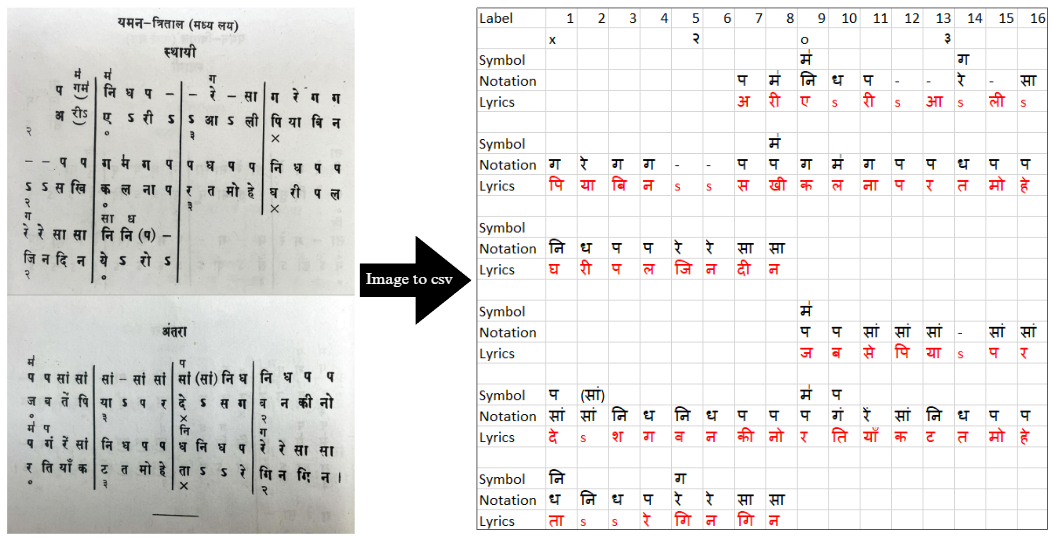}
    \caption{The composition \textit{yeri aali} as printed in Kramik Pustak Malika \cite{Bhatkhande} (left); machine-readable CSV file for the two verses of 2 lines each (right)}
    \label{fig:canonical_csv}
\end{figure}

 We convert the canonical notation of the \textit{bandish}, as available in the Bhatkhande book, into a machine-readable CSV format, retaining the note label (both pitch and lyric syllable) and timing information \cite{madhumitha}. 
 As seen in Figure~\ref{fig:canonical_csv}, the notes and lyric syllables are placed on a \textit{tala} grid, containing as many columns as the number of beats in the \textit{tala}.  In this study, all \textit{bandish} are set in \textit{teentaal} (16-beat cycle), with salient beats like the \textit{sam} (downbeat) and \textit{khali} explicitly marked. Each line of the \textit{bandish} spans 16 beats, with each beat containing either a syllable (or, rarely, multiple syllables), a rest (empty string), or a continuation of the previous note (‘s’). Each cycle is described with three rows: the lyrics, ornamentation symbols (if any), and the corresponding notes (\textit{sargam}). Additionally, the row below the lyrics indicates \textit{vibhag} symbols, marking the start of each quarter cycle.

\section{Signal and Text Processing Methods}

Given that computational methods can help musicological analyses achieve scale, we investigate audio processing algorithms for the annotation discussed in the previous section. 
Automatic speech recognition models are of limited use on singing voice due to the significant differences in acoustic parameters, coupled with the compromised quality of vocals obtained with automatic source separation networks. Forced alignment of a singing voice with lyrics continues to be a topic of research and is especially challenging in non-English language settings such as ours \cite{vaglio2020multilingual}.
We restrict ourselves to the automatic detection of syllable onsets for the present, followed by text based alignment of the canonical lyrics with the manually labeled audio syllables.  



\subsection{Onset detection pipeline}
We explore methods that exploit the prominent spectrum transitions that mark onsets. The consonant to vowel boundary is marked by a significant increase of energy in certain frequency bands. The band energies are found to take on low values for semi-vowels, nasals and voiced stops \cite{subbandonsetdet}. Apart from the temporal change in band energies, we try the differencing of MFCC (mel frequency cepstral coefficients) features due to their ability to capture phone identity in a compact representation. All differences are computed using a smoothened derivative function across audio frames of 10 ms durations \cite{biphasicfilter}. The resulting novelty function is examined for local maxima.




\begin{figure*}[t]
    \centering
    \includegraphics[width=\textwidth]{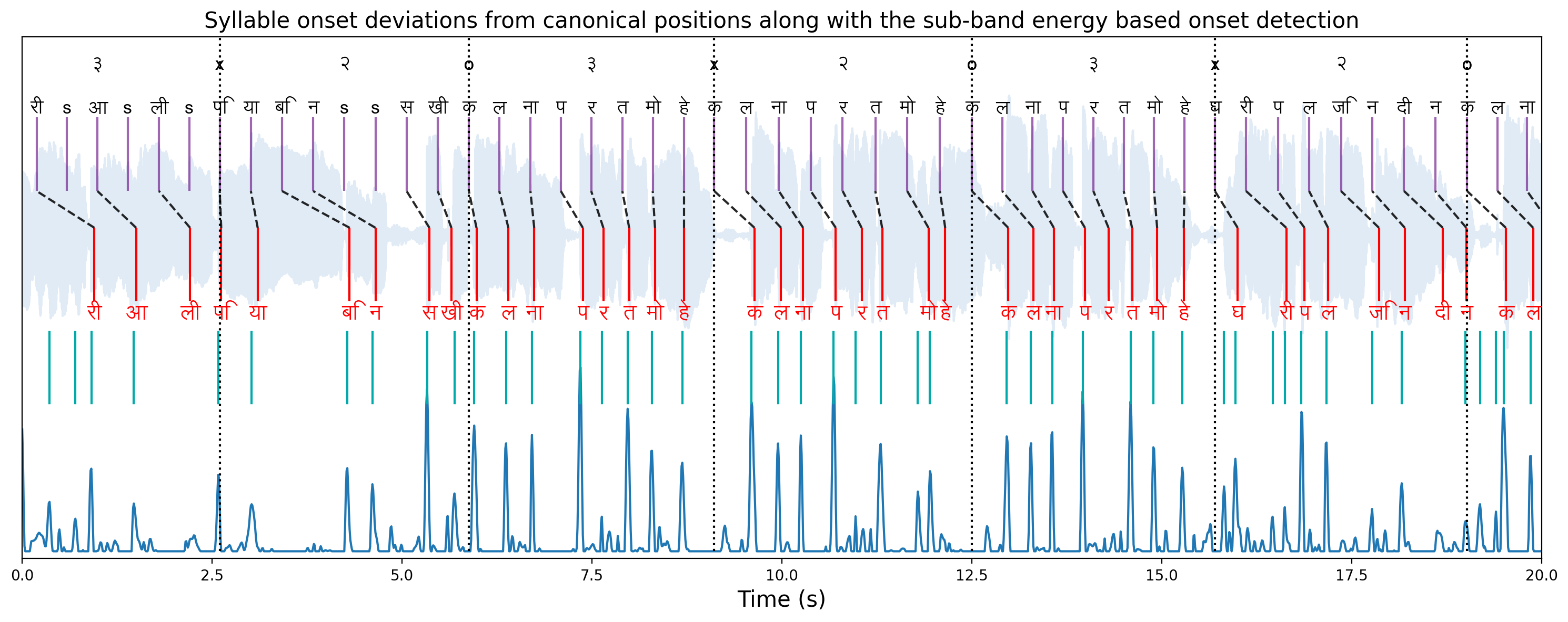}
    \caption{Syllable alignment with canonical beat positions (purple) and detected onsets (blue) from novelty function (bottom most curve) for an excerpt of  \textit{yeri aali} by Ashwini Bhide Deshpande. The canonical lyric syllables aligned with their corresponding matra appear right on top (with vibhag marked in vertical dotted lines). In red font, are the audio-time stamped syllables. Note the timing variations across 3 repetitions of the same line over the duration 6 s -15 s.}
    \label{fig:devn_nov_combined}
\end{figure*}

Table ~\ref{tab:onset_eval} compares the detection performance of the different acoustic features on our manually labelled dataset of 2041 syllable onsets across 8 singers in terms of onset detection precision and recall with a true positive defined as a detected onset occurring within 50 ms of a manually annotated onset. The sub-band energy-based model (Model 1) outperforms the difference MFCCs-based model (Model 2) overall. An analysis of the missed detections revealed that for CV transitions involving liquids (r, l), semi-vowels (w,y), voiceless obstruent (h) and nasals (n, m) the energy is observed not to drop as much in the expected sub-band (640-2800 Hz). 


\vspace{-2pt}
\begin{table}[h!]
\centering
\footnotesize
\begin{tabular}{lccc}
\toprule
Model & Precision & Recall & F1 score \\
& (\%) & (\%) & (\%) \\
\midrule
\multicolumn{4}{c}{For maximized F1 score} \\
\midrule
Difference MFCCs        & 73.8 & 73.6 & 73.7 \\
Sub-band energy    & 82.5 & 77.9 & 80.1 \\
\midrule
\multicolumn{4}{c}{For a fixed recall} \\
\midrule
Difference MFCCs        & 62.1 & 80.0 & 70.2 \\
Sub-band energy    & 79.5 & 80.0 & 79.8 \\
\bottomrule
\end{tabular}
\vspace{3pt}
\caption{Onset detection performance across 2041 syllables of 14 audio recordings by 8 singers}
\label{tab:onset_eval}
\end{table}

Another contributing factor is the audio quality that depends on the age of the recordings and the fact that they are concert performances that were later source-separated. 
Finally, the ornamentation and fluid movement of the pitch Hindustani classical music triggers changes within spectral bands across frames and gives rise to false positives in onset detection. These challenges need to be addressed using learning from the data, if possible. The scarcity of labeled data may be offset in future by using self-supervised learning or available pretrained audio models.


\subsection{Text alignment and syllable mapping}

We describe here the required lyric alignment of the audio recording at the syllable level. Given the two sequences, canonical notation 
and the manually labeled syllables with the audio time-stamps, we need to establish the correspondence between each canonical syllable and its audio realization. With this then, we can compute the timing deviation of a sung syllable with reference to its canonical beat position. The first step is to choose a relevant large interval for the mapping, and we selected the half cycle of the \textit{teentaal} rhythmic cycle (beats 1-8 and 9-16) for this purpose. We use text-level alignment by considering the canonical sequence corresponding to one-half cycle of beats and searching for the best-matched subsequence using a sliding window across the labeled sequence obtained from the audio. 



Next, a stage of refinement is carried out where individual syllables within the mapped interval were aligned by iterating over the manually marked intervals and canonical intervals, ensuring correct mapping even in cases of missing or extra syllables, or shifted syllables from neighbouring intervals. An example of the resulting alignment appears in Figure ~\ref{fig:devn_nov_combined}. An immediate observation is the varying time lag of the audio-realised (manually labeled in red) syllables with reference to their canonical locations (in black, as derived from the composition text). The achieved mapping enables us to deduce systematic relations, if any, between the two sequences of syllables, also related to musical note events. 


\section{Observations and Discussion}

\begin{figure}
    \centering
    \includegraphics[width=1\linewidth]{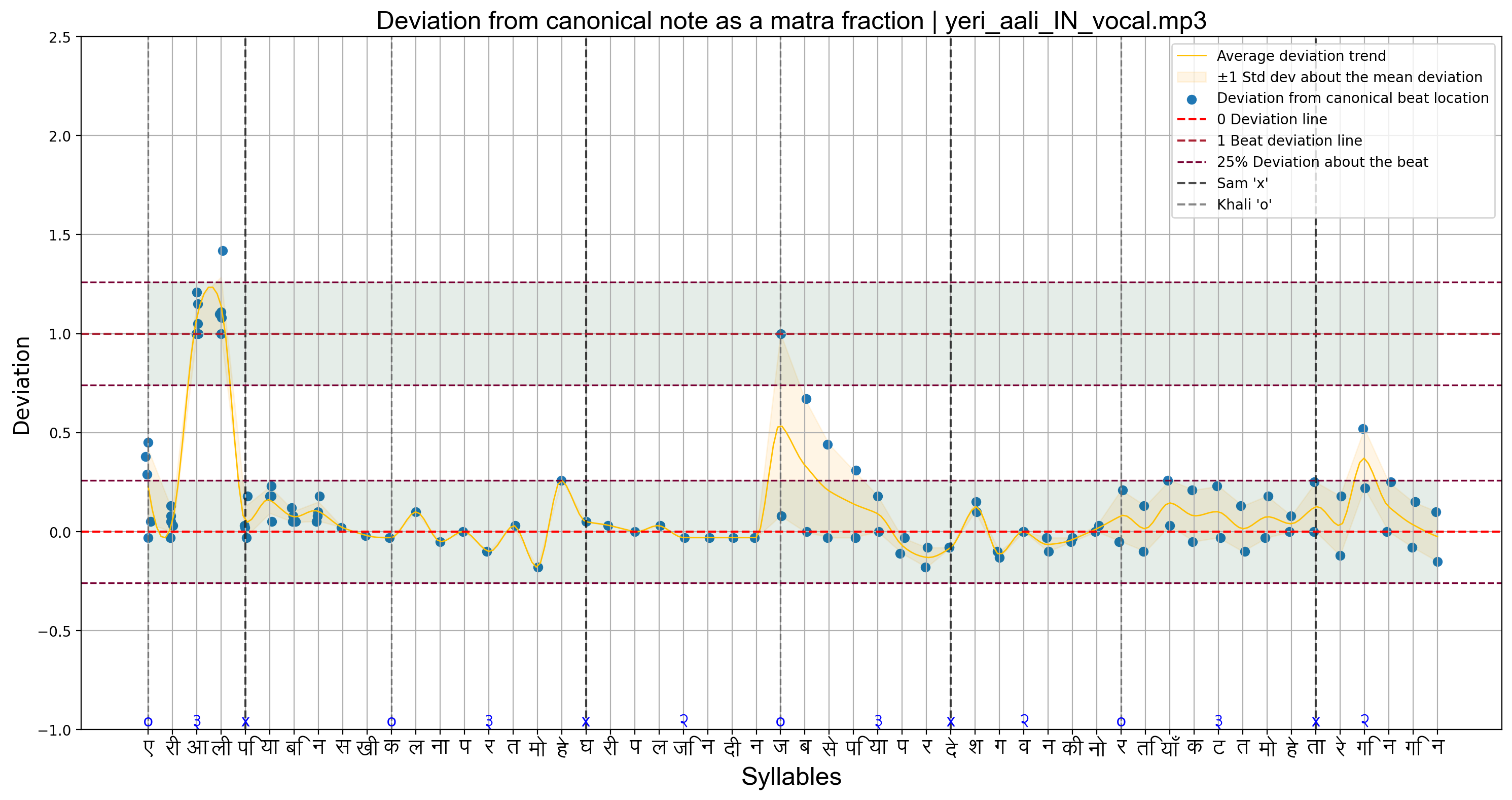}
    \caption{Deviation of the sung syllable onsets from the canonical locations as a fraction of beat duration for \textit{yeri\_aali\_IN} across the four \textit{bandish} lines}
    \label{fig:devn_yeri_aali_IN}
\end{figure}

\begin{figure}
    \centering
    \includegraphics[width=1\linewidth]{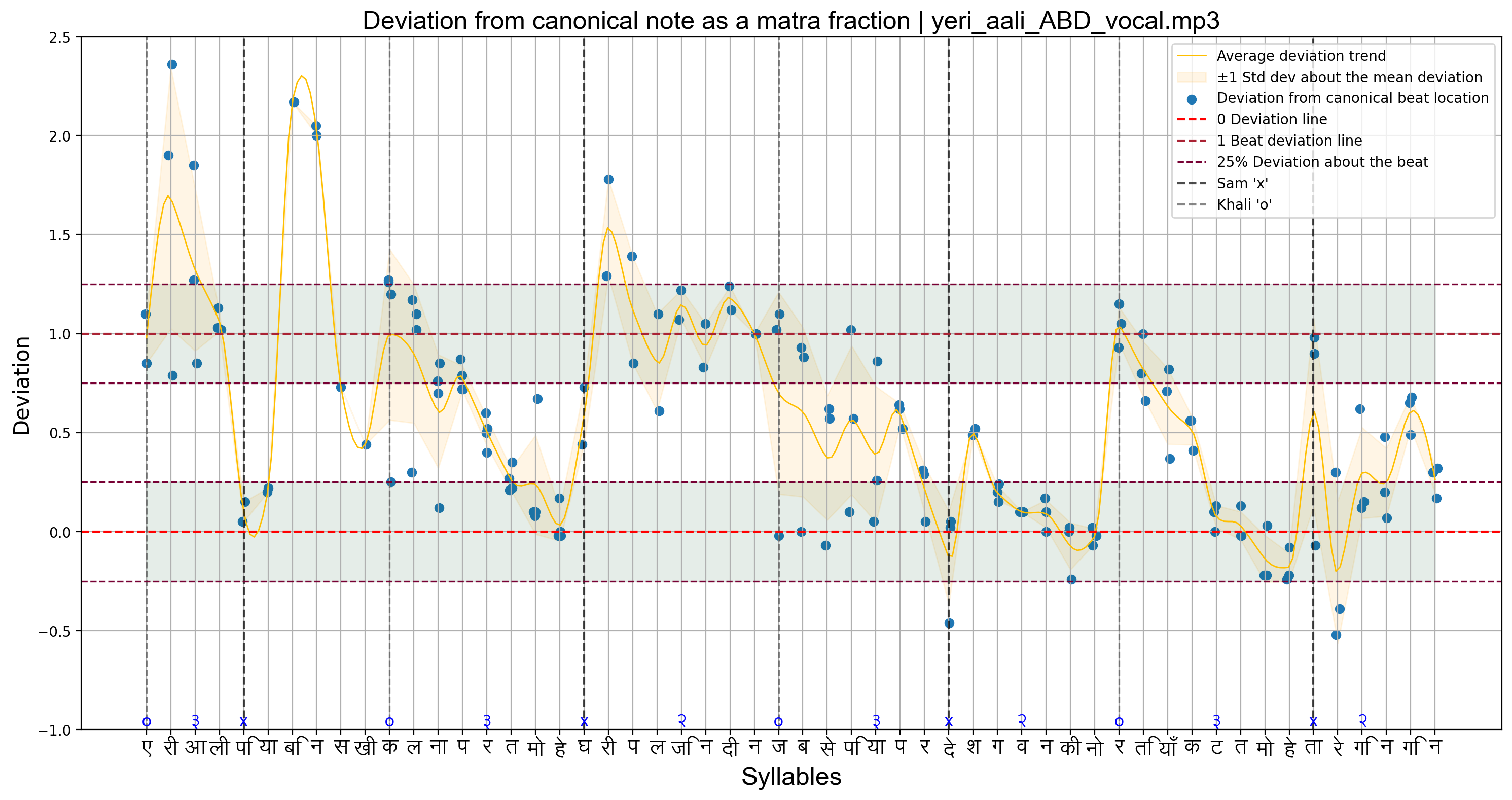}
    \caption{Deviation of the sung syllable onsets from the canonical locations as a fraction of beat duration for \textit{yeri\_aali\_ABD} across the four \textit{bandish} lines}
    
    \label{fig:devn_yeri_aali_ABD}
\end{figure}

We now present a comparative study of two performances of the same \textit{bandish} by two different singers in our dataset. One performance, as self-reported by the artist, Pt. Nirody (IN), is a close reproduction of the canonical notation. We assume that this influences the rhythmic aspect and we expect the note events to largely coincide with the beat positions indicated in the canonical notation. The second performance by a prominent artist, Ashwini Bhide-Deshpande (ABD), is typical of the concert setting and therefore strongly expected to display variations including expressive timing. We present a comparative analysis of the recordings via the Figures ~\ref{fig:devn_yeri_aali_IN} and ~\ref{fig:devn_yeri_aali_ABD} where the temporal dynamics are visualized in different ways.

The downward trend of the points on the plot from the start of the cycle towards the cycle boundary indicates that the singer starts a \textit{bandish} line a bit late (lagging) in a rather time-free manner, and compensates for this delay by compressing the following notes ensuring that a lyric segment - whether spanning a full cycle, half or even quarter, is not taken across cycle boundaries. This is essential for keeping up with the rhythm as well as providing a sense of resolution. Consistent with this strategy, the syllables corresponding to the later beats of a cycle, and the \textit{sam} of the new cycle, tend to show far less temporal deviation. A cluster of points in the band near fractional deviation equal to one full \textit{matra}, and very few instances within the 0 deviation band indicates a recurrent structural modification that the artist employs in rendering the composition. Generally, this kind of structural deviation is limited to one \textit{matra}, and in fewer cases 2 \textit{matras}.

A strong trend across performances in our dataset is for the singers to be mostly lagging 
as observed in Figures ~\ref{fig:devn_nov_combined}, ~\ref{fig:devn_yeri_aali_ABD}, and still complete the line within the particular cycle or half-cycle by compressing the syllables together, i.e. shortening the gap between consecutive syllables, as well as using up the empty beats or note extensions ('s' markings) according to the canonical notation. It may be remarked that this bears a striking similarity to tempo rubato in jazz \cite{ashley}.

\begin{figure}
    \centering
    \includegraphics[width=1\linewidth]{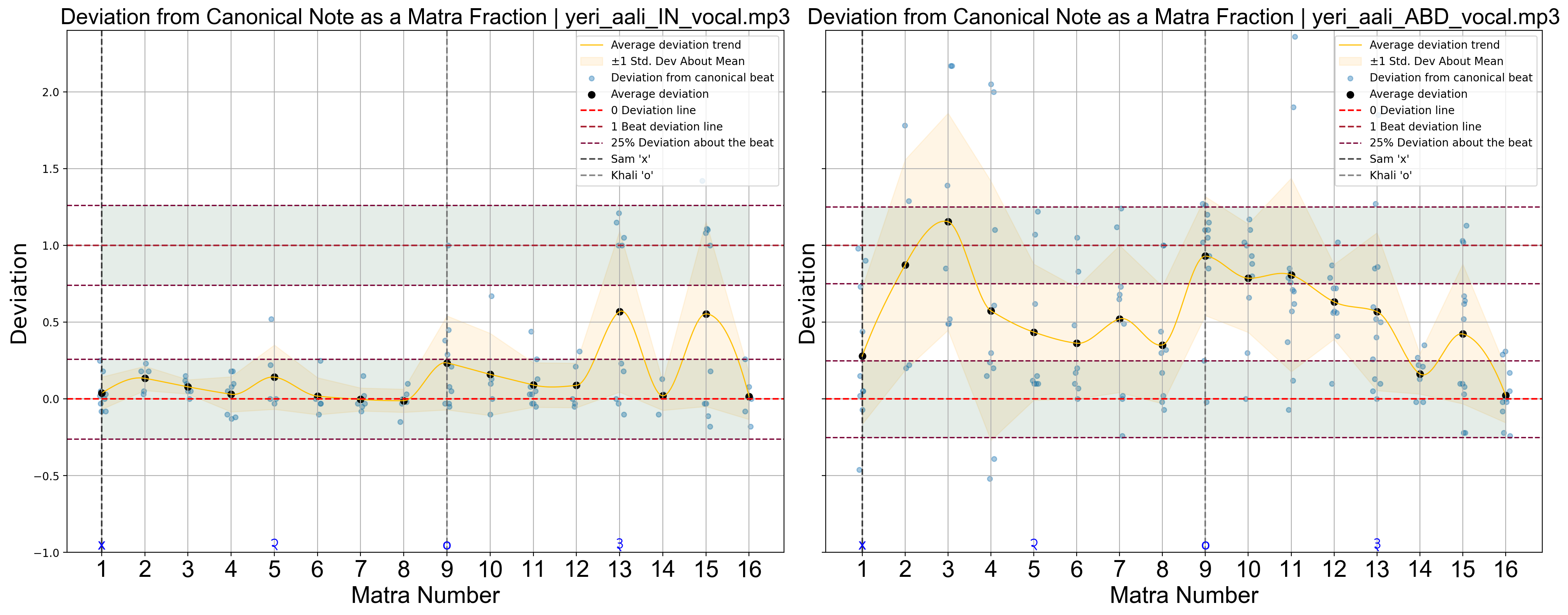}
    \caption{Deviation of actual syllable onsets from the canonical locations as a fraction of the beat across the 16 beats of a cycle, averaged over 9 cycles for \textit{yeri\_aali\_IN} and over 10 cycles for \textit{yeri\_aali\_ABD}}
    \label{fig:compressed_beat_devn}
\end{figure}

Figure ~\ref{fig:compressed_beat_devn} collapses multiples rhythm cycles to achieve something like a 'fingerprint' of artist's expressive timing where the extent of timing flexibility at the beat level clearly depends on its location in the cycle. As expected from the previous discussion, a significant drop in the timing deviation is observed on the matra just preceding the \textit{sam} and the \textit{khali}. Further, the contrast between the two performances is clearly manifested in the timing deviation pattern, validating our choice of representation for timing expressiveness.

\section{Conclusion and Future Work}

This study investigates how Hindustani vocalists strategically navigate the tension between adhering to the metric structure and injecting expressive timing variations by analyzing the alignment of syllable onsets with the rhythmic cycle. The results demonstrate how this approach can reveal distinctive patterns in the temporal phrasing of different artists, providing insights into the role of expressive timing in the aesthetics of this musical tradition. Through the audio analysis of performances of a popular composition, we have demonstrated how expert musicians take liberties with timing, particularly in regions that fall away from primary structural markers, such as the downbeat and the regions with low syllable density. In these sections, the performer may deviate from the strict tempo, introducing variations that contribute to the individuality and expressiveness of their rendition. We conclude that temporal dynamics contribute significantly towards the expressiveness of a rendition, and there is a manner to it. 

Furthermore, our study introduces a systematic pipeline designed to measure certain temporal dynamics. With future improvements in the automatic audio processing pipeline including lyric alignment, large-scale analyses of such distinctive stylistic elements can become possible. It also offers valuable inputs for generative modelling. Through this framework, the model can learn to emulate the nuanced temporal flexibility found in traditional performances, thus enabling the generation of music that retains both the structure and expressive timing characteristic of a particular artist or genre. This research could pave the way for more authentic, style-specific generative AI models for music, capable of reproducing complex temporal dynamics that showcase the expressiveness and stylistic identity of traditional music renditions.


\bibliographystyle{IEEEtran}
\bibliography{Expressive_Timing_in_Hindustani_Vocal_Music}

\end{document}